\documentclass[aps,pre,twocolumn,superscriptaddress,nofootinbib,longbibliography]{revtex4-2}
 
\usepackage{amsmath,amssymb,bm,mathtools}
\usepackage{graphicx}
\usepackage{xcolor,comment}

\usepackage{hyperref}
\hypersetup{
    colorlinks=true,
    linkcolor=blue,  
    citecolor=red,   
    urlcolor=red   
}

\makeatletter
\def\Xint#1{\mathchoice
    {\XXint\displaystyle\textstyle{#1}}%
    {\XXint\textstyle\scriptstyle{#1}}%
    {\XXint\scriptstyle\scriptscriptstyle{#1}}%
    {\XXint\scriptscriptstyle\scriptscriptstyle{#1}}%
    \!\int}
\def\XXint#1#2#3{{\setbox0=\hbox{$#1{#2#3}{\int}$}
    \vcenter{\hbox{$#2#3$}}\kern-.5\wd0}}
\def\ddashint{\Xint=}
\makeatother
 
\newcommand{\pss}{\mathbb{P}_{\text{ss}}}

\begin{document}
 
\title{Freezing transition and
shock-like transients in ballistic random walks under threshold resetting
}
 
\author{Rahul Das}
\email{rahuldas@imsc.res.in}
\affiliation{The Institute of Mathematical Sciences, CIT Campus, Taramani,
Chennai 600113, India}
\affiliation{Homi Bhabha National Institute, Training School Complex,
Anushakti Nagar, Mumbai 400094, India}

\author{Satya N Majumdar}
\email{satyanarayan.majumdar@cnrs.fr}
\affiliation{Laboratoire de Physique Théorique et Modèles Statistiques (LPTMS),
Université de Paris-Sud, Bâtiment 100, 91405 Orsay Cedex, France}
 
\author{Arnab Pal}
\email{arnabpal@imsc.res.in}
\affiliation{The Institute of Mathematical Sciences, CIT Campus, Taramani,
Chennai 600113, India}
\affiliation{Homi Bhabha National Institute, Training School Complex,
Anushakti Nagar, Mumbai 400094, India}
\date{\today}
\begin{abstract}
We investigate the transient dynamics of a one-dimensional ballistic random walker under threshold resetting (TR). The walker starts at $x_0>0$, moves towards a threshold at the origin with a velocity drawn from a distribution $\phi(v)$, and on reaching the threshold is reset to $x_0$ with a new sampled velocity. Whenever the mean first-passage time to the threshold is finite, the walker reaches a steady state that is uniform on $[0,x_0]$ and independent of $\phi(v)$. The relaxation towards it, in contrast, is highly sensitive to the behavior of $\phi(v)$. For a large class of $\phi(v)$-s, we find, rather remarkably, that the relaxation rate depends explicitly on the spatial coordinate, in contrast to conventional relaxation phenomena, where the rate is typically position-independent. Furthermore, this space-dependent relaxation rate eventually freezes to a constant, position-independent value beyond a critical position, a phenomenon we identify as \textit{freezing transition}. Additionally, for a power-law velocity distribution, the system exhibits an anomalously slow, algebraic relaxation towards the steady state. We also showed that a sharp cutoff in $\phi(v)$ generates a shock-like discontinuity in the particle propagator that travels towards the threshold, with a jump height that decays with time. These results demonstrate how velocity fluctuations, despite leaving the steady state unchanged, can profoundly influence the transient spatial structures and showcase rich non-equilibrium behavior. Our analytical results are corroborated by Monte Carlo simulations, showing an excellent agreement.
\end{abstract}
 
\maketitle

\section{Introduction}\label{sec:intro}
Stochastic resetting, whereby an ongoing stochastic process is interrupted and returned to a prescribed configuration, has emerged as an important paradigm in statistical physics since its introduction in the context of diffusion \cite{evans_diffusion_2011,evans2011diffusion}. Despite its conceptual simplicity, resetting can profoundly alter the dynamics of a system, breaking detailed balance and generating non-equilibrium steady states (NESS) \cite{evans_diffusion_2011,evans2011diffusion,gupta2014fluctuating,pal2015diffusion,pal2016diffusion,eule2016non,nagar2016diffusion, de2026dynamically}, as well as rich transient phenomena, including anomalous relaxation \cite{majumdar2015dynamical,pal2016diffusion,singh2020resetting}. Beyond its role in shaping non-equilibrium dynamics, resetting has proved to be a powerful strategy for optimizing first-passage processes, offering improvements in both search efficiency and reliability \cite{evans_diffusion_2011,evans2011diffusion,evans2013optimal,pal2016diffusion,pal_first_2017,reuveni_optimal_2016,kusmierz2014first,pal2023thermodynamic,pal2019landau,evans2018effects,pal2019first,sar2023resetting}. The scope of resetting has since expanded well beyond its original setting in diffusion, finding applications in rather unrelated topics such as chemical reaction kinetics \cite{reuveni2014role,robin2018single,biswas2023rate}, biophysical search processes \cite{roldan2016stochastic,budnar2019anillin,rajoria2026broad,kondev2026biological}, operations research \cite{bonomo2021mitigating,roy2024queues}, and even in economics \cite{stojkoski2021geometric,jolakoski2023first,stojkoski2022income}. In parallel, resetting protocols have been realized experimentally using optical traps, enabling the investigation of a variety of non-equilibrium phenomena \cite{tal2020experimental,besga2020optimal,faisant2021optimal,goerlich2023experimental,kundu2025emulating,biroli2026experimental,goerlich2026time,sherf2026stochastic}, and have also been implemented using mechanical robots \cite{paramanick2024uncovering,paramanick2024programming}. For broader perspectives on the development and emerging directions of the field, we refer the reader to the reviews \cite{evans_stochastic_2020,pal2022inspection,gupta2022stochastic,pal2024random,nagar2023stochastic}.

In the conventional paradigm of stochastic resetting, reset events occur at random times determined by an external stochastic process (say, an external timer), independently of the instantaneous state of the system. However, many natural and engineered systems instead
reset themselves when an internal variable crosses a critical value. An
integrate-and-fire neuron discharges and returns to its resting potential once
the membrane potential reaches a firing threshold~\cite{burkitt2006review}; a
stop-loss order liquidates a position when a price limit is
breached~\cite{shiryaev2007optimal}; a loaded fibre bundle sheds a fibre once
its rupture threshold is exceeded~\cite{hansen2015fiber}; a cell divides once
a critical amount of some internal variable has
accumulated~\cite{corigliano2026essential,kumar2026branching}. 
This event-driven protocol, in which the reset is triggered
by the system itself is referred to as
\textit{threshold resetting} (TR)~\cite{biswas2025target, biswas2026optimal,das2026non}, and is also known as \textit{first-passage resetting} in~\cite{de2020optimization,de2021optimization,biroli2026first}. Naturally, the reset events are, by construction, correlated with the trajectory that
produces them.

Threshold resetting remains a relatively nascent area of research, with only a handful of studies addressing its effects in single- and many-particle systems. The existing literature has explored a few key aspects of this protocol across different settings. For instance,
De Bruyne \textit{et al} analyzed the
steady state in a finite system and the reset-reward cost objective function of a single diffusing particle under TR \cite{de2020optimization,de2021optimization}. 
Biroli \textit{et al} computed the exact steady-state joint position distribution for $N>2$ Brownian searchers with threshold resetting, showed evidence of dynamically emergent correlations in the steady state and, moreover, computed the exact statistics of several physical observables in this steady state analytically, including the extreme value statistics, the gap statistics and the full-counting statistics \cite{biroli2026first}.
On the other hand, Biswas \textit{et al} carried out a first-passage study involving $N$ diffusing and ballistic particles subjected to TR and found non-trivial optimization behavior in the mean search time and further showed that optimal resetting can forestall larger cost  \cite{biswas2025target,biswas2026optimal}. 
García-Valladares \textit{et al} studied first-passage-time optimization of a single one-dimensional Brownian searcher to distributed targets in the presence of both threshold resetting and Poissonian resetting \cite{garcia2026target}.  
Recent work by Chen and Bressloff on bounded two-dimensional domains with small interior targets has shown that the effectiveness of boundary resetting depends sensitively on the choice of reset position, the target configuration, and the waiting time at the boundary~\cite{chen20262d}. Frimpong \textit{et al} studied position distribution, first-passage- and occupation- time statistics of a sluggish random walker (discrete and continuous) in one dimension under threshold resetting \cite{frimpong2026sluggish}, generalizing earlier works on sluggish walker with Poissonian resetting~\cite{del2026proxitaxis} and with uniform memory resetting~\cite{boyer2026ultra}. 

\begin{figure}[t]
    \centering
    \includegraphics[width=1\linewidth]{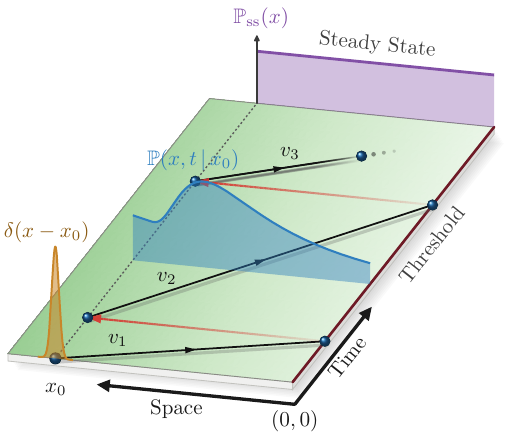}
    \caption{
    Schematic illustration of a ballistic particle under threshold resetting (TR). The particle starts from an initial position $x = x_0$ and moves toward the threshold at the origin with a random velocity drawn from a velocity distribution $\phi(v)$. As soon as it reaches the threshold, it resets back to the initial position and renews its dynamics. Repeated resetting may lead to a steady-state distribution $\mathbb{P}_{\text{ss}}(x)$ in the long-time limit. In this system, the particle always remains confined in the spatial regime $x \in [0,x_0]$.}
    \label{fig:schematic}
\end{figure}

In a preceding work, we investigated non-equilibrium dynamics of a general stochastic process subjected to TR \cite{das2026non}. We 
derived a general criterion for the emergence of a NESS, showing that whenever the mean first-passage
time (MFPT) to the threshold is finite, the steady state is the ratio of the
mean local time (MLT) to the MFPT. To illustrate this result, we analyzed a drift-diffusion process under TR, computed its NESS and further examined its transient behavior. The aim of the current work is to examine non-equilibrium dynamics of a different stochastic process namely a ballistic walker with randomized velocities. 
As illustrated schematically in Fig.~\ref{fig:schematic}, the walker starts at $x_0>0$ and moves ballistically towards the threshold at the origin with a velocity drawn from a prescribed distribution $\phi(v)$. Upon reaching the threshold, the walker is instantaneously returned to its initial position, where a new velocity is independently sampled from the same distribution $\phi(v)$. A distinctive feature of this model is that the motion remains entirely deterministic between successive resetting events, with stochasticity entering solely through the random selection of velocities at the beginning of each excursion. We would like to understand how these velocity fluctuations lead to a non-equilibrium steady state and the relaxation mechanism towards stationarity.

We now summarize the key results of our work. Whenever the MFPT to the threshold is finite, TR drives the ballistic walker into a universal steady state, uniform on $[0,x_0]$ and independent of the velocity distribution $\phi(v)$, whereas the approach to it depends strongly on the low-velocity behavior of $\phi(v)$. Most strikingly, the relaxation rate itself can depend on position. 
This is unlike in most systems where the relaxation rate is a constant that does not depend on space variables.
This space-dependent rate arises when $\phi(v)\sim v^{-(m+1)}e^{-a/v}$ as $v\to0$ ($m\in\mathbb{N}$). Intriguingly, the rate can undergo a \emph{freezing transition}, i.e., it depends on position up to a critical point $x^\star$ and then freezes onto a constant plateau, thus becoming space-independent. We derive a necessary and sufficient condition on $\phi(v)$ to observe this transition. In addition, a power law $\phi(v)\sim v^{\alpha}$ ($0<\alpha<1$) as $v\to0$ instead yields a slow $t^{-\alpha}$ relaxation. We also find that a sharp cutoff $v_c$ (lower or upper) in the support of $\phi(v)$ generates a shock-like discontinuity in the particle propagator, travelling towards the threshold at speed $v_c$. Its jump height at the discontinuity decays exactly as $\phi(v_c)/t$, so the shock directly measures the velocity distribution at its cutoff. In a strict deterministic limit $\phi(v)=\delta(v-v_0)$, the system is found to attain a periodic steady state. Overall, velocity fluctuations leave rich and striking signatures in the transient dynamics while preserving the universality of the steady state.

For the readers' convenience, we now provide the structure of the paper. 
Section~\ref{sec:framework} recalls the
renewal framework for TR and the representation of the steady state as a ratio
of the MLT to the MFPT. Section~\ref{sec:SSball} introduces the
ballistic model, derives its propagator in Laplace space, and establishes the
universal uniform steady state. The transients are taken up next: the freezing
transition in Sec.~\ref{sec:FreezTrans}, the power-law relaxation in
Sec.~\ref{sec:powerlaw}, and the shock-like fronts in Sec.~\ref{sec:shock}.
Section~\ref{sec:deltaFunc} treats the deterministic limit, and we conclude in
Sec.~\ref{sec:conclude}. Supporting derivations are collected in the
Appendices.

\section{General framework: threshold resetting}
\label{sec:framework}
 
We consider a particle moving in one dimension that starts at $x= x_0$ and
evolves under some intrinsic stochastic dynamics. Whenever the particle
reaches the origin $x=0$ (which we designate as the \textit{threshold}) it is
instantaneously reset to its initial position $x=x_0$, after which it resumes
its dynamics anew. 
The central object of interest is the probability density $\mathbb{P}(x,t|x_0)$ of
finding the particle at position $x$ at time $t$ starting from initial position $x_0$.

The renewal structure of TR was introduced in Ref. \cite{de2020optimization} and was developed in Ref. \cite{das2026non} into a general characterization of the steady state for an arbitrary underlying dynamics. We recall below only those elements that are needed for the ballistic problem, and refer the reader to Ref. \cite{das2026non} for the detailed derivations.
 
\subsection{The renewal structure}
 
Because each reset erases the memory of the trajectory and restarts it from
$x_0$, TR is a renewal process, and the propagator obeys the renewal
equation~\cite{de2020optimization,biroli2026first,das2026non}
\begin{equation}
\mathbb{P}(x,t|x_0)=\mathbb{G}_+(x,t|x_0)+\int_0^t d\tau\,\mathbb{F}(\tau|x_0)\,\mathbb{P}(x,t-\tau|x_0),
\label{eq:renewal}
\end{equation}
where $\mathbb{G}_{+}(x, t |x_0)$ is the propagator in the presence of an absorbing boundary at the origin (here, $+$ stands as an indicator that the particle stays in the positive half-line) and $\mathbb{F}(t|x_0)$ is the first-passage time (FPT) density
to the threshold. The two terms have a transparent meaning. The first accounts
for trajectories that have not reset at all in $[0,t]$: the threshold acts as an
absorbing boundary because a particle that reaches it stops and is reset,
never continuing past. The second term accounts for trajectories that have
reset at least once: the first reset occurs in $[\tau,\tau+d\tau]$ with
probability $d\tau\,\mathbb{F}(\tau|x_0)$, after which the particle restarts from $x_0$
and propagates for the remaining time $t-\tau$.
 
A Laplace transform, $\widetilde{f}(s)=\int_0^\infty dt\,e^{-st}f(t)$, converts the
convolution in Eq.~\eqref{eq:renewal} into an algebraic relation
\begin{equation}
\widetilde{\mathbb{P}}(x,s|x_0)=\frac{\widetilde{\mathbb{G}}_+(x,s|x_0)}{1-\widetilde{\mathbb{F}}(s|x_0)},
\label{eq:renewalLaplace}
\end{equation}
The two inputs are not independent. Integrating the constrained propagator
over the final position gives the survival probability
$\mathbb{S}(t|x_0)=\int_0^\infty dx\,\mathbb{G}_+(x,t|x_0)$, and the FPT density follows
as $\mathbb{F}(t|x_0)=-\partial_t \mathbb{S}(t|x_0)$; using $\mathbb{S}(0|x_0)=1$, this
yields~\cite{das2026non}
\begin{equation}
    \widetilde{\mathbb{F}}(s|x_0)= 1- s \int_0^{\infty} dx \, \widetilde{\mathbb{G}}_+ (x,s|x_0).
\end{equation}
Consequently, Eq. \eqref{eq:renewalLaplace} becomes
\begin{align}\label{eq:PropSimplified}
    \widetilde{\mathbb{P}}(x,s|x_0)=\frac{\widetilde{\mathbb{G}}_+(x,s|x_0)}{s \int_0^\infty dx~ \widetilde{\mathbb{G}}_+(x,s |x_0)}.
\end{align}
Hence the full
resetting dynamics is fixed entirely by the reset-free constraint propagator
$\widetilde{\mathbb{G}}_+(x, s|x_0)$. For a wide class of underlying processes, this is
known in closed form, which makes the propagator under TR analytically
accessible.
 
\subsection{Steady state under TR}
\label{sec:ss}

If a steady-state (SS) exists, $\mathbb{P}(x, t|x_0)$ becomes time-independent at long times, i.e., $\mathbb{P}(x, t \to \infty|x_0) = \mathbb{P}_{\text{ss}}(x)$. Consequently, its Laplace transform with respect to $t$ behaves for small $s$ as
\begin{align}
\widetilde{\mathbb{P}}(x,s|x_0)=& \int_0^{\infty} \mathbb{P}(x,t|x_0) e^{-st} dt  \quad \\ &\to \frac{1}{s} \mathbb{P}_{\text{ss}}(x)  ~~\text{as}~ s\to 0. \nonumber
\end{align}
This gives rise to the following relation (also known as the final value theorem) \cite{biroli2026first,das2026non}
\begin{align}\label{eq:SSlimit}
    \mathbb{P}_{\text{ss}}(x) &= \lim_{s \to 0} s \widetilde{\mathbb{P}}(x, s|x_0) =  \frac{\widetilde{\mathbb{G}}_{+}(x, s =  0|x_0)}{\int_0^\infty dx~ \widetilde{\mathbb{G}}_+(x,s = 0 |x_0)}.
\end{align}
When the mean first-passage time (MFPT) $\langle T\rangle$ to the threshold is
finite, equation~\eqref{eq:SSlimit}
then collapses to the compact form derived in \cite{das2026non}
\begin{equation}
\mathbb{P}_{\rm ss}(x)=\frac{\langle L(x|x_0)\rangle}{\langle T\rangle},
\label{eq:ness-ratio}
\end{equation}
where the quantities in the above equation can be computed from the reset-free propagator
\begin{align}
&\langle T\rangle =\int_0^\infty dt\,t\,\mathbb{F}(t|x_0) = \int_0^\infty dx \int_0^\infty dt~ \mathbb{G}_+(x, t|x_0),
\label{eq:mfpt}\\[4pt]
&\langle L(x|x_0)\rangle =\int_0^\infty dt\,\mathbb{G}_+(x,t|x_0).
\label{eq:mlt}
\end{align}
Here $\langle L(x|x_0)\rangle$ is the mean local time (MLT), i.e.\ the mean
time the particle spends at position $x$ during a single excursion between two
successive threshold crossings; it follows by averaging
$L(x|x_0)=\int_0^\infty dt\,\delta[X(t)-x]$ over the absorbing-boundary
ensemble~\cite{das2026non}. Eq.~\eqref{eq:ness-ratio} therefore states
that $\mathbb{P}_{\rm ss}(x)$ is the average time spent at $x$ in one cycle
divided by the average duration of a cycle, a cycle being the interval between
two threshold-driven first-passage events.

Two consequences of Eq.~\eqref{eq:ness-ratio} are worth mentioning. First,
Eqs.~\eqref{eq:mfpt} and \eqref{eq:mlt} imply $\langle T\rangle=\int_0^\infty
dx\,\langle L(x|x_0)\rangle$, so that $\mathbb{P}_{\rm ss}(x)$ is
automatically normalized to unity whenever it exists. Second, this ratio form
constrains the admissible shape of the steady state: a finite
$\langle T\rangle$ forces the MLT either to be finite for every $x$, giving a
continuous $\mathbb{P}_{\rm ss}(x)$, or to develop integrable singularities at
isolated points, giving a steady state that diverges there while remaining
normalizable~\cite{das2026non}. The ballistic model studied below falls in
the first class, and in fact realizes an interesting version of it: the MLT is
not merely finite but independent of $x$ throughout the accessible region,
which is the origin of the universal steady state obtained in
Sec.~\ref{sec:uniSS}.

A finite MFPT thus provides a sufficient criterion for the existence of a NESS
in a noisy system. Interestingly, when the underlying
dynamics is deterministic, successive renewal cycles never dephase, and the long-time state is strictly
periodic rather than stationary even though $\langle T\rangle$ is finite. We
return to this case in Sec.~\ref{sec:deltaFunc}; in
Secs.~\ref{sec:SSball}-\ref{sec:shock} the randomness of the velocity
supplies the dephasing between cycles and Eq.~\eqref{eq:ness-ratio} applies.

\section{Ballistic Particle under TR: Model and universal steady state}\label{sec:SSball}

We now specialize the underlying process to purely ballistic motion. A particle
is released at $x=x_0$ and moves toward the origin at a constant velocity $v$
drawn from a probability density $\phi(v)$. When it reaches the threshold at
the origin, it is reset to $x_0$, and a new velocity is sampled from
$\phi(v)$; the cycle then repeats (see Fig. \ref{fig:schematic}). We take $v\in[0,\infty)$, i.e.\ every
excursion is directed toward the threshold. Restricting to $v \ge 0$ guarantees that each excursion terminates at the 
threshold. Otherwise, the particle escapes to infinity, TR fails to confine 
it, and no non-equilibrium steady state exists.

Other simple systems include many particles with simultaneous resetting after hitting the threshold or a single particle with a drift towards the threshold, ensuring finite MFPT and consequently a steady-state under TR, which are studied in Ref. \cite{biswas2025target,biswas2026optimal,biroli2026first, das2026non}.

\subsection{Reset-free propagator and first-passage density}

During a single excursion starting from the initial position $x = x_0$, the particle travels deterministically, i.e., $x(t)=x_0-vt$ 
until it hits the origin at time $x_0/v$. The propagator with an absorbing boundary at the origin is therefore
\begin{align}
    \mathbb{G}_+(x,t|x_0)=\int_0^\infty dv\,\phi(v)\,\delta\!\big(x-(x_0-vt)\big)\nonumber \\
    \times\Theta\!\Big(\tfrac{x_0}{v}-t\Big)\Theta(x_0-x),
\label{eq:Gtime}
\end{align}
where the Heaviside functions confine the particle to be between $x = 0$ and $x = x_0$. 
As the propagator under TR can be computed exactly in Laplace space following Eq.\eqref{eq:PropSimplified}, we need to compute the Laplace transform of the above quantity, which reads
\begin{align}
    \widetilde{ \mathbb{G}}_+(x,s|x_0) =\Theta(x_0-x)\int_0^\infty \frac{dv\,\phi(v)}{v}\,
    \exp\!\Big[-\frac{s(x_0-x)}{v}\Big].
    \label{eq:Glap}
\end{align}
The full expression of the propagator under TR in Laplace space immediately follows [Eq.\eqref{eq:PropSimplified}]
\begin{equation}
    \widetilde{\mathbb{P}}(x,s|x_0)=\Theta(x_0-x)\,
    \frac{\displaystyle\int_0^\infty \frac{\phi(v)}{v}\,e^{-s(x_0-x)/v}\,dv}
    {\displaystyle 1-\int_0^\infty \phi(v)\,e^{-sx_0/v}\,dv}.
\label{eq:Plap}
\end{equation}
Now we are going to rewrite Eq. \eqref{eq:Plap} with a change of variable $u = 1/v$, which will be useful in Sec. \ref{sec:FreezTrans} and Sec. \ref{sec:powerlaw}. After the change in variable, the Laplace space propagator can be re-expressed as
\begin{equation}\label{eq:mapSing}
\widetilde{\mathbb{P}}(x,s|x_0)=-\Theta(x_0 - x)\frac{\widetilde{\psi}^\prime\big(s(x_0-x)\big)}
{1-\widetilde{\psi}(sx_0)},
\end{equation}
where $\widetilde{\psi}(p) = \int_0^\infty du~ e^{-pu}~\psi(u)$ is the Laplace transform of $\psi(u)$ and $\psi(u)$ is defined in terms of $\phi$ as follows
\begin{equation}
\psi(u)=\frac{1}{u^{2}}\,\phi\!\Big(\frac{1}{u}\Big),
\label{eq:chipsi}
\end{equation}
also $\widetilde{\psi}^\prime(p) = d\widetilde{\psi}/dp$.
Consequently, $\widetilde{\psi}^\prime(p)$ and $\widetilde{\psi}(p)$ are singular at the same
points in the complex-$p$ plane, although the nature of the singularity differs: a pole of
$\widetilde{\psi}$ of order $m$ is a pole of $\widetilde{\psi}^\prime$ of order $m+1$, and a branch
point of one is a branch point of the other. This fact will be quite useful when we discuss the relaxation properties later.

\subsection{Universal steady state: uniform distribution}\label{sec:uniSS}

By substituting the result from Eq. \eqref{eq:Gtime} into Eq. \eqref{eq:mfpt} and Eq. \eqref{eq:mlt}, we obtain the MFPT and MLT for a ballistic particle with a general velocity distribution $\phi(v)$,
\begin{align}
    \langle T \rangle&=x_0\int_0^\infty\frac{\phi(v)}{v}\,dv,\label{eq:MFPT_ball}\\
    \langle L(x|x_0)\rangle &=\Theta(x_0-x)\int_0^\infty\frac{\phi(v)}{v}\,dv.\label{eq:MLT_ball}
\end{align}
Both are controlled by the same integral $\int_0^\infty dv\,\phi(v)/v$, which
converges only if $\phi(v)/v$ is integrable near $v=0$; that is, $\phi(v)$ must
vanish sufficiently fast as $v\to0$. Physically, this is a statement about slow
excursions: if there is a non-negligible chance of drawing a near-zero
velocity, the corresponding excursion takes an anomalously long time; consequently, the MFPT to find the threshold diverges. Provided
$\phi(v)\to0$ fast enough that the integral converges, the ratio in
Eq.~\eqref{eq:ness-ratio} is finite and, strikingly, independent of the shape of $\phi(v)$:
\begin{equation}
\;\pss(x)=\frac{1}{x_0}\,\Theta(x_0-x).\;
\label{eq:uniform}
\end{equation}
The stationary state is a universal uniform distribution on
$[0,x_0]$. This is a direct consequence of Eq.~\eqref{eq:MFPT_ball},\eqref{eq:MLT_ball} and Eq. \eqref{eq:ness-ratio}: for ballistic
motion, the local time is the same at every point along an excursion, so
the particle is, on average, equally likely to be found anywhere between the
reset point and the threshold, regardless of how the speed is distributed.

The uniformity of the steady state admits a natural geometric interpretation through a change in geometry. By reversing the direction of motion and identifying the two endpoints of the interval, the original dynamics can be mapped onto a ballistic random walker moving around a circle of circumference \(2\pi\). The angular velocity \(\omega\) is independently drawn from a distribution \(\rho(\omega)\) after each completed revolution. During each cycle, the walker traverses the entire circle, spending a time \(\omega^{-1}d\theta\) in an angular interval \(d\theta\). Consequently, the mean residence time per unit angle is \(\langle\omega^{-1}\rangle\), while the mean cycle duration is \(2\pi\langle\omega^{-1}\rangle\). The stationary angular density therefore follows from the ratio of the mean residence time per unit angle to the mean cycle duration
\begin{equation}
P_{\mathrm{ss}}(\theta)
=\frac{\langle\omega^{-1}\rangle}
{2\pi\langle\omega^{-1}\rangle}
=\frac{1}{2\pi},
\qquad 0\leq\theta<2\pi,
\end{equation}
provided the mean cycle duration is finite and the renewal process admits a stationary state. This geometric construction is related in spirit to the idea of equidistribution, as exemplified in Weyl's theorem \cite{weyl1916gleichverteilung,kuipers1974uniform}, and the result is a simple consequence of ergodicity via Weyl's theorem. The geometric mapping thus provides an intuitive interpretation of the uniform steady state while the distribution of cycle durations can still govern the relaxation towards this universal stationary density.

\section{Freezing transition}\label{sec:FreezTrans}
While the steady state is universal, the approach to it is not: the detailed shape of $\phi(v)$ leaves a strong imprint on the transient dynamics. Depending on $\phi(v)$, we get different types of relaxation behaviors, e.g., exponential, power-law, etc. In this section, we discuss only exponential relaxations, specifically when the slowest relaxation rate depends on particle position $x \in [0, x_0]$. We find that there exists a freezing transition in the relaxation rate which, quite interestingly, changes from an independent to space-dependent form, governed by the properties of $\phi(v)$. Such a freezing transition was identified earlier in \cite{sabhapandit2020freezing} in the escape rate of an overdamped diffusing particle from a trap. 

\subsection{An exact case}\label{sec:solvableExample}
\begin{figure*}
    \centering
    \includegraphics[width=1\linewidth]{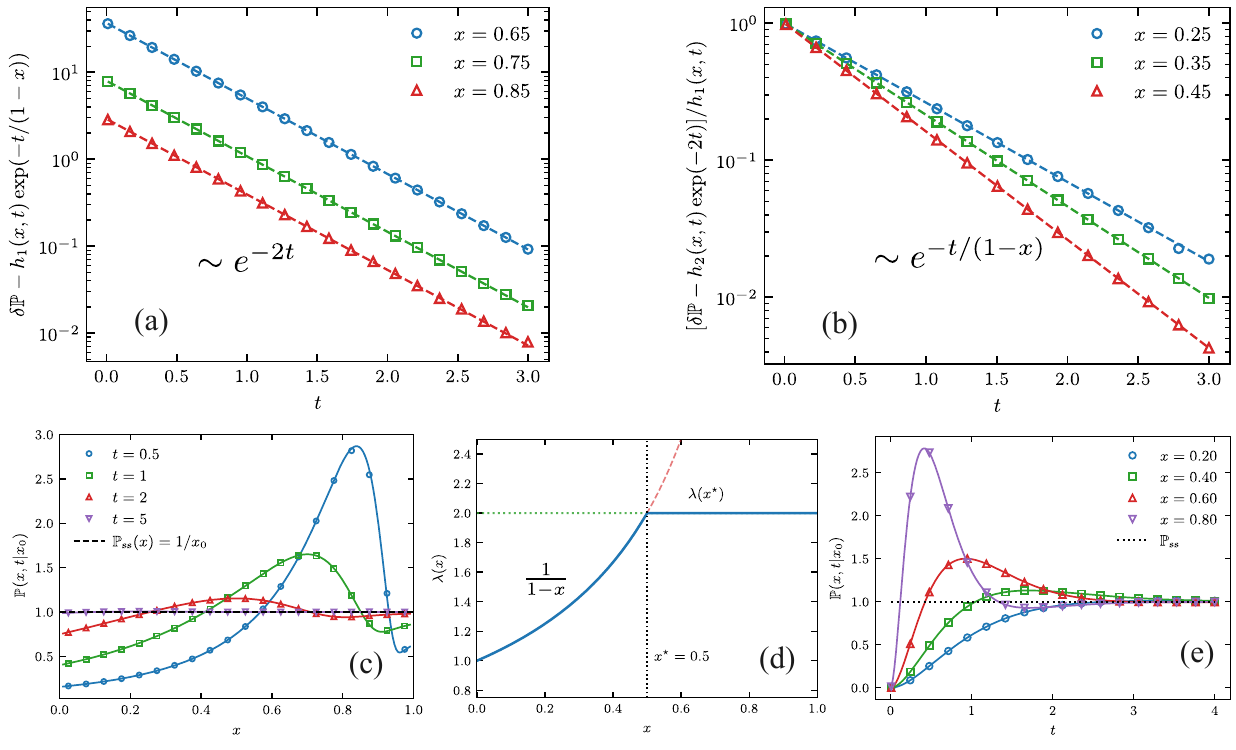}
    \caption{A comprehensive plot for $n = 3$ case, i.e., $\phi(v) \propto v^{-3}e^{-v_0/v}$ case (see appendix~\ref{app:n=3 case} and Sec. \ref{sec:solvableExample}) and its verification with simulation results. The freezing transition happens at $x^\star = x_0/2$ point [Eq.\eqref{eq:xstar-values}]. There exist two relaxation rates: $\lambda(x) = v_0/(x_0 - x), 2v_0/x_0$ by Eq. \eqref{eq:lambdaX}, one of which is space-dependent and the other is not. Panel (a) shows the space-independent relaxation for different spatial points $x > x^\star$, whereas Panel (b) shows the space-dependent relaxation for spatial points $x < x^\star$. We here defined $\delta \mathbb{P} = \mathbb{P}(x, t|x_0) - \mathbb{P}_{\text{ss}}(x)$. The dashed lines are exact theoretical results, and the markers are simulation results, where the propagator $\mathbb{P}(x, t|x_0)$ is generated through Monte Carlo simulations. Panel (c) shows the propagator behaviour with respect to $x$ at different time points $ t$, and how it reaches the steady state $\mathbb{P}_{\text{ss}}(x) = 1/x_0$ at long times. Panel (d) shows the relaxation rate $\lambda(x)$ as a function of space $x$, which increases as $v_0/(x_0 - x)$ until $x = x^\star$, then freezes at $\lambda(x^\star)$ [Eq. \eqref{eq:lambdaX}]. Panel (e) presents the non-monotonic relaxation of different spatial points [exact expression given in Eq. \eqref{eq:Pexact}]. Parameters: $x_0 = 1, v_0 = 1$.}
    \label{fig:n=3 case}
\end{figure*}

To expose the physics of the relaxation properly, we adopt the velocity distribution of a one-parameter family,
\begin{equation}
\phi(v)=\frac{v_0^{\,n-1}}{\Gamma(n-1)\,v^{n}}\,e^{-v_0/v},\qquad n>1
\label{eq:phin}
\end{equation}
which is normalised and has a convergent $\int_0^\infty dv\,\phi(v)/v$ for every
$n>1$, so the universal steady state~\eqref{eq:uniform} always holds. What distinguishes the members of the
family is the singularity structure of the propagator in Laplace space for different $n$ values, and hence the relaxation. Substituting Eq.~\eqref{eq:phin} into Eqs.~\eqref{eq:Glap} gives,
\begin{align}
\widetilde{\mathbb{G}}_+(x,s|x_0)&=\frac{(n-1)v_0^{\,n-1}}{\big(v_0+s(x_0-x)\big)^{n}}\Theta(x_0 -x),
\label{eq:Gn}
\end{align}
and, through Eq.~\eqref{eq:PropSimplified},
\begin{equation}
\widetilde{\mathbb{P}}(x,s|x_0)=\frac{(n-1)v_0^{\,n-1}}{\big(v_0+s(x_0-x)\big)^{n}}\,
\frac{\Theta(x_0- x)}{1-\big(v_0/(v_0+sx_0)\big)^{n-1}}.
\label{eq:Pn}
\end{equation}
The explicit finite-time propagator, obtained by
inverse Laplace transform of Eq. \eqref{eq:Pn}, is given in Appendix~\ref{app:finite}. The spatial behavior of the propagator for $n = 3$ at different times is shown in Fig. \ref{fig:n=3 case}(c); non-monotonic relaxations of different spatial points to the steady state are shown in Fig. \ref{fig:n=3 case}(e).

To study the steady state and relaxation behavior, it is quite important to analyze the singularity structure of the propagator in Eq. \eqref{eq:Pn}.
Depending on whether $n$ is an integer or non-integer, this structure changes, giving rise to different relaxation behavior. Both cases are discussed below.

\begin{figure}[htbp]
    \centering
    \includegraphics[width=1\linewidth]{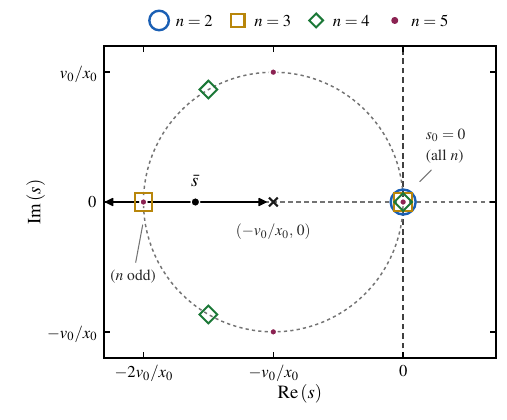}
    \caption{The pole positions of $\bar{s}$ and $s_k$ for $n = 2, 3, 4, 5$ [from Eq.~\eqref{eq:sbar} and \eqref{eq:sk}]. As we increase the value of n, more poles from $s_k$ appear and lie on the circle centered at $(-v_0/x_0, 0)$. There is always a pole at the origin, which is responsible for the steady state in the long-time limit. $n \ge 4$ gives rise to complex conjugate poles. There also exists a pole at $s = (-2v_0/x_0, 0)$ only for odd $n$ values.}
    \label{fig:sk}
\end{figure}

\medskip
\noindent\textbf{\emph{Integer n case}:}
Due to the pole present in the constrained propagator in Laplace space $\widetilde{\mathbb{G}}_+(x, s|x_0)$ given in Eq. \eqref{eq:Gn}, we find a space-dependent $n-$th order pole in the TR propagator in Eq. \eqref{eq:Pn} at
\begin{equation}
\bar{s}=-\frac{v_0}{x_0-x},
\label{eq:sbar}
\end{equation}
which sweeps along the line $\bar{s} \in (-\infty, -v_0/x_0)$ depending on the position of the particle $x \in [0, x_0]$. The other denominator in $\widetilde{\mathbb{P}}$, involving the Laplace-space first-passage (FP) density, gives rise to space-independent simple poles
\begin{equation}
s_k=\frac{v_0}{x_0}\Big[\cos\theta_k-1+i\sin\theta_k\Big],
\quad \theta_k=\frac{2\pi k}{n-1},
\label{eq:sk}
\end{equation}
with $k=0,1,\dots,n-2$. These lie on a circle of radius $v_0/x_0$ in the complex-$s$ plane centered at
$(-v_0/x_0,0)$ that touches the origin; the $k=0$ pole at $s=0$ produces the
steady state, while $n\ge4$ introduces complex-conjugate pairs and hence damped
oscillatory relaxation. The positioning of the poles in the complex plane is shown in Fig.~\ref{fig:sk}.

The rightmost non-zero singularity dictates the slowest relaxation mode, and the rate is defined by the absolute real part of its position in the complex-$s$ plane. A freezing transition occurs when the space-dependent singularity [Eq.~\eqref{eq:sbar}] crosses the rightmost non-zero space-independent singularity [Eq.~\eqref{eq:sk}, $k=1$] at a spatial point $x = x^\star$. Equating their real parts gives the exact position of the crossing, namely the freezing point
\begin{equation}
x^\star=x_0\left[1-\frac{1}{1-\cos\!\big(2\pi/(n-1)\big)}\right],
\label{eq:xstar-n}
\end{equation}
which depends on $n$ but not on $v_0$. Explicitly,
\begin{equation}
x^\star=
\begin{cases}
x_0/2, & n=3,\\
x_0/3, & n=4,\\
0, & n=5,\\
<0\ (\text{not physical}), & n\ge6 .
\end{cases}
\label{eq:xstar-values}
\end{equation}
For $n=2$ there is no space-independent pole other than the one at the origin,
so the relaxation rate is space-dependent everywhere,
$\lambda(x)= |\bar{s}| =v_0/(x_0-x)$, and no freezing transition occurs. For $n\ge5$ the
crossing has moved out of the domain $[0, x_0]$, and the relaxation is space-independent
everywhere. A genuine freezing transition with a position-dependent regime
giving way to a plateau inside $(0,x_0)$ exists only for $n=3$ and $n=4$. The behavior of the relaxation rate $\lambda(x)$ for these cases is as follows:
\begin{align}\label{eq:lambdaX}
    \lambda(x) = 
    \begin{cases}
        \dfrac{v_0}{x_0 - x}, & x < x^\star, \\[0.7pc]
        \dfrac{v_0}{x_0}\left[1 - \cos\left(\dfrac{2\pi}{n -1}\right)\right], & x \ge x^\star.
    \end{cases}
\end{align}
Fig. \ref{fig:n=3 case}(d) plots the behaviour of $\lambda(x)$ for $n = 3$, whereas the simulation verification of these relaxations for different spatial points is shown in Fig. \ref{fig:n=3 case}(a), (b).

\medskip
\noindent\textbf{\emph{Non-integer n case}:} 
We do not observe any freezing transition for non-integer $n$ values. The relaxation rate $\lambda$ is space-independent for all $x \in [0, x_0]$. This is evident from the singularity structure of $\widetilde{\mathbb{P}}(x, s|x_0)$, given in Eq. \eqref{eq:Pn}. There exists-
\begin{enumerate}
    \item a space-independent branch point at $s_{b1} = -v_0/x_0$,
    \item a space-dependent branch point at $s_{b2} = -v_0/(x_0 -x)$,
    \item poles arising from the roots of the denominator $1 - (v_0/(v_0 + sx_0))^{n-1} = 0$.
\end{enumerate}
Notice that $s_{b2}$ moves for varying $x$, but is bounded on the right by $s_{b1}$ as $x \to 0$. There would always be a pole at $s = 0$, giving rise to a steady state, whereas other poles may or may not exist on the principal Riemann sheet depending on the specific value of $n$ [Appendix \ref{app:nonIn}].
As the rightmost singularity (except the $s = 0$ one) is always space-independent for any $x \in [0, x_0]$ in this case, we do not observe any freezing transition in the relaxation rate $\lambda$.

\subsection{General theory of the space-dependent rate and the freezing transition}\label{sec:freezGeneral}

To obtain the general criterion on the velocity distribution $\phi(v)$ responsible for a freezing
transition in the relaxation rate, we examine the rightmost singularity of $\widetilde{\mathbb{P}}(x,s|x_0)$ in the
complex-$s$ plane, other than $s = 0$, which is responsible for the steady state. For this discussion, we are going to use the expression in Eq. \eqref{eq:mapSing}. A freezing transition occurs when the space-dependent singularity crosses the rightmost space-independent (non-zero) one at a spatial point $x^\star \in (0, x_0)$.

\medskip
\noindent\textbf{\emph{Space-dependent singularity}:} Since $\psi(u)\ge 0$, the rightmost singularity of its Laplace transform
$\widetilde{\psi}(p)$ lies on the real axis \cite{widder1941}; call it $p=-a$ with $a>0$. Consequently, its derivative $\widetilde{\psi}^\prime(p)$ also has a singularity there.
Through
$s(x_0-x)=-a$, the singularity of the propagator $\widetilde{\mathbb{P}}(x, s|x_0)$ by Eq. \eqref{eq:mapSing} is located at
\begin{equation}
    s_{\text{num}}=- \frac{a}{x_0 -x }.
\end{equation}
This singularity lies on the real axis $s \in (-\infty, -a/x_0]$ for any $x \in [0, x_0]$ in the complex-$s$ plane. This singularity in $\widetilde{\mathbb{P}}(x, s|x_0)$ makes the relaxation rate space-dependent. The singularity at $s_{\text{num}}$ must be a pole for a freezing transition to exist, as discussed in the next subsection.

\medskip
\noindent\textbf{\emph{Space-independent singularity}:} The denominator of
Eq.~\eqref{eq:mapSing} contributes in two distinct ways. Each of them contributes to the space-independent singularities.

\emph{\underline{(i) Zeros of $1-\widetilde{\psi}(sx_0)$:}} Normalization
$\int_0^\infty du\,\psi(u)=1$, guarantees the trivial root $s=0$, which is the pole responsible for
the steady state. The maximum-valued (real part) non-trivial root gives rise to the rightmost space-independent pole(s) in $\widetilde{\mathbb{P}}(x, s|x_0)$, whose position on the real axis of the complex-$s$ plane is as follows
\begin{equation}
    s_{\text{den}} = -\frac{b}{x_0},
\end{equation}
where the root is defined as
\begin{align}\label{eq:b_sing}
    b &=  - \operatorname{max}\left[\operatorname{Re} p : \widetilde{\psi}(p) = 1, p \ne 0 \right], \nonumber\\
      &= - \operatorname{max}\left[\operatorname{Re} p: \int_0^\infty dv~\phi(v) \exp({- p/v}) = 1, p \ne 0 \right].
\end{align}
This depends on the full velocity distribution $\phi(v)$.

\emph{\underline{(ii) The rightmost singularity of $\widetilde{\psi}$ itself is at $sx_0=-a$:}} This
survives in $\widetilde{\mathbb{P}}$ only if it is a branch point. If it is a pole, then
$1-\widetilde{\psi}\to\infty$ there and Eq.~\eqref{eq:mapSing} develops a zero rather
than a singularity. 

If there exists a branch point at $s = -a/x_0$ in $\widetilde{\mathbb{P}}(x, s|x_0)$, it bounds the space-dependent singularity $s_{\text{num}} = -a/(x_0 - x)$ on the right side of the complex-$s$ plane so that they can't cross each other for any $x \in [0, x_0]$. Consequently, there does not exist any freezing transition. This is the same reason we did not find a freezing transition for the non-integer $n$ case in Sec. \ref{sec:solvableExample}. This indicates that the rightmost singularity of $\widetilde{\psi}(p)$ has to be a pole, then large-$u$ behavior of $\psi(u)$ is as follows
\begin{equation}\label{eq:psiHigh}
    \psi(u) \simeq \mathcal{C} u^{m - 1} e^{-a u}, \qquad m \in \mathbb{N},
\end{equation}
where $m$ denotes the order of the pole ($m = 1$ is for a simple pole). Note: this condition fixes the nature of the space-dependent singularity $s_{\text{num}} = -a/(x_0 - x)$ discussed earlier as a pole of order $m+1$, as it arises due to $\widetilde{\psi}^\prime$. Returning to the
velocity variable $v=1/u$, Eq. \eqref{eq:psiHigh} sets a condition on the low-velocity {\color{blue}behavior} of $\phi(v)$,
\begin{equation}\label{eq:low-v-tail}
\phi(v)\simeq \mathcal{C} v^{- (m + 1)}e^{-a/v},\qquad v\to 0,
\end{equation}
where $m \in \mathbb{N}$.
This is a hard constraint on $\phi(v)$ in the small-$v$ limit, acting as a necessary condition for the existence of a freezing transition, but it is not sufficient. This property in $\phi(v)$ ensures the existence of a space-dependent relaxation mode, as we have already seen for the example used in Sec. \ref{sec:solvableExample} for the integer-$n$ case -- there, the representative case $\phi(v) \propto v^{-n} e^{-v_0/v}$ from Eq. \eqref{eq:phin} connects to Eq. \eqref{eq:low-v-tail} through the relation $n = m + 1$.

\medskip
\noindent\textbf{\emph{Condition for a freezing transition}:}
A freezing transition occurs when $s_{\text{num}}$ crosses $s_{\text{den}}$ at the transition point $x = x^\star$, i.e.,
\begin{equation}
-\frac{a}{x_0-x^\star}=-\frac{b}{x_0}
\implies
x^\star=x_0\Big(1-\frac{a}{b}\Big).
\label{eq:xstar-general}
\end{equation}
For $x<x^\star$, the space-dependent singularity $s_{\text{num}}$ dictates the relaxation; for $x>x^\star$, it shifts to the space-independent one $s_{\text{den}}$. A genuine transition therefore lies inside the domain,
$x^\star\in(0,x_0)$, if and only if
\begin{equation}\label{eq:ab}
0<a<b,
\end{equation}
where $a$ is set by the essential singularity in low-$v$ {\color{blue}behavior} of $\phi(v)$ as shown in Eq. \eqref{eq:low-v-tail} and $b$ depends on the full distribution $\phi(v)$ through Eq. \eqref{eq:b_sing}. Eq. \eqref{eq:low-v-tail} and Eq. \eqref{eq:ab} together constitute the necessary and sufficient condition for the existence of a freezing transition in the ballistic threshold resetting model.

\section{Power-law relaxation}\label{sec:powerlaw}

The freezing transition analysis in Sec. \ref{sec:FreezTrans} assumed an essential singularity $\phi(v) \sim e^{- a/v}$ in the $v \to 0$ limit, giving rise to exponential relaxation. A slower approach to the steady state occurs when the low-velocity $(v \to 0)$ behavior is a power law instead, such as
\begin{equation}\label{eq:plv}
    \phi(v) \simeq A v^\alpha, \quad \text{with} \quad \alpha > 0.
\end{equation}
The first-passage (FP) time density inherits the low-velocity behavior of $\phi$ directly,
\begin{align}
    &\mathbb{F}(t|x_0) = \int_0^\infty dv~\phi(v) \delta\left(t - \frac{x_0}{v}\right) =\frac{x_0}{t^{2}}\,\phi\!\left(\frac{x_0}{t}\right),\\
    &\mathbb{F}(t |x_0) \simeq A\,x_0^{1+\alpha}\,t^{-(2+\alpha)}, \qquad \mathrm{as} ~~t \to \infty,
\end{align}
showing a heavy-tail behavior. For $0 < \alpha < 1$, the second and higher moments of the FP time density diverge, but the MFPT to find the threshold at the origin is finite, and the universal steady state of Eq. \eqref{eq:uniform} exists. For $1 < \alpha < 2$, the first two moments of the FP density are finite, and so on (narrow density). We focus here only on the broad FP time density regime, i.e., $0 < \alpha < 1$.

\begin{figure}[t]
    \centering
    \includegraphics[width=1\linewidth]{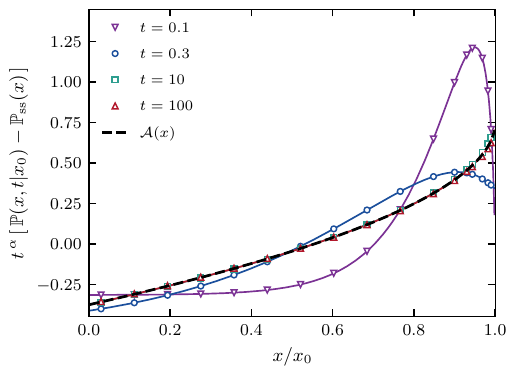}
    \caption{Power-law relaxation towards the steady state for $\phi(v) = v^\alpha e^{-v/v_0}/[v_0^{\alpha + 1}\Gamma(\alpha+1)]$ with $\alpha = 0.5$. Consequently, the mean FP time is $\langle T\rangle = x_0/(\alpha v_0)$ for this velocity distribution. This plot shows how the spatial behavior of $t^\alpha \delta\mathbb{P}$ collapses to $\mathcal{A}(x)$ at long times, justifying Eq. \eqref{eq:PLrelxn}. Markers denote the simulation results and solid lines denote theory obtained using numerical inverse Laplace transform of the exact Laplace space propagator given by Eq. \eqref{eq:Plap}.  Parameters used: $x_0 = 1, v_0 = 1$.}
    \label{fig:powerLaw}
\end{figure}

The relaxation behavior can be studied using the small-$s$ (or large $t$) behavior of the Laplace Space propagator $\widetilde{\mathbb{P}}(x, s|x_0)$. In this section, we are again going to use the form given in Eq. \eqref{eq:mapSing} in terms of the Laplace transform of $\psi(u)$. Recall that $u = 1/v$ is the changed variable mentioned. The small-$v$ behavior in Eq.~\eqref{eq:plv} translates into algebraic large-$u$ tail of the the function
\begin{equation}
    \psi(u)= \frac{1}{u^{2}}\,\phi\!\Big(\frac{1}{u}\Big) \simeq A\,u^{-(2+\alpha)},\qquad u\to\infty.
\end{equation}
The corresponding Laplace space expansion of this function and its derivative in the small-$s$ limit reads
\begin{align}
    &\widetilde{\psi}(sx_0) = 1  - sx_0 \langle u\rangle + A\Gamma(-1-\alpha)(sx_0)^{\alpha+1} + o(s^{\alpha + 1}),\\
    &\widetilde{\psi}^\prime(s(x_0 - x)) = -\langle u\rangle - A\Gamma(-\alpha)s^\alpha (x_0 - x)^\alpha + o(s^\alpha),
\end{align}
where $\langle u \rangle = \int_0^\infty u~ \psi(u) du = \int_0^\infty \phi(v)/v~ dv = \langle 1/v \rangle$ and we have used the property of gamma function that $\Gamma(-\alpha) = -(\alpha + 1)\Gamma(-\alpha - 1)$. By substituting the above results into Eq. \eqref{eq:mapSing}, we find the small-$s$ behavior of the propagator
\begin{align}
\widetilde{\mathbb{P}}(x,s|x_0)
&= \Theta(x_0-x)
\bigg[
\frac{1}{x_0s}
+ \frac{A}{\langle T\rangle}
\bigg\{
\Gamma(-\alpha)(x_0-x)^\alpha
\nonumber\\
&
+ \Gamma(-\alpha-1)x_0^\alpha
\bigg\}
s^{\alpha-1}
+ o\!\left(s^{\alpha-1}\right)
\bigg].
\end{align}
where we have used the MFPT definition $\langle T\rangle = x_0\langle 1/v\rangle$. The first term corresponds to the steady state $\mathbb{P}_{\text{ss}}(x)$, and the second term contributes to the power-law relaxation. After the Laplace inversion, the propagator in the long-time limit behaves as follows (Tauberian Theorem)
\begin{align}\label{eq:PLrelxn}
    t^\alpha~ \delta\mathbb{P} &\simeq \frac{A}{\alpha \langle T\rangle}\left[\frac{x_0^\alpha}{1 + \alpha} - (x_0 - x)^\alpha \right]\Theta(x_0 - x),\nonumber\\
    &= \mathcal{A}(x),
\end{align}
where we have defined $\delta \mathbb{P} = \mathbb{P}(x, t|x_0) - \mathbb{P}_{\text{ss}}(x)$. The amplitude satisfies $\int_0^{x_0}dx \mathcal{A}(x) = 0$, since $\mathbb{P}(x, t|x_0)$ and $\mathbb{P}_{\text{ss}}(x)$ are separately normalized in the domain $[0, x_0]$. Simulation verification of the above result is shown in Fig. \ref{fig:powerLaw} for $\phi(v) = v^\alpha e^{-v/v_0}/[v_0^{\alpha + 1}\Gamma(\alpha+1)]$ with $0 < \alpha < 1$. The physical reason behind this slow relaxation is the abundance of slow excursions with $v \to 0$ that populate arbitrarily long first-passage
times (even though their mean stays finite), and it is this heavy supply of
slow modes that governs the late
stage of relaxation. 

\section{Shock-like transients}\label{sec:shock}

A qualitatively different transient arises when the velocity distribution has sharp bounds, e.g., $\phi(v)$ defined in $v \in [v_c, \infty)$.
Such a cutoff $v_c$ produces moving spatial discontinuities, i.e., shock-like fronts in the propagator $\mathbb{P}(x, t|x_0)$.

\begin{figure}[t]
    \centering
    \includegraphics[width=1\linewidth]{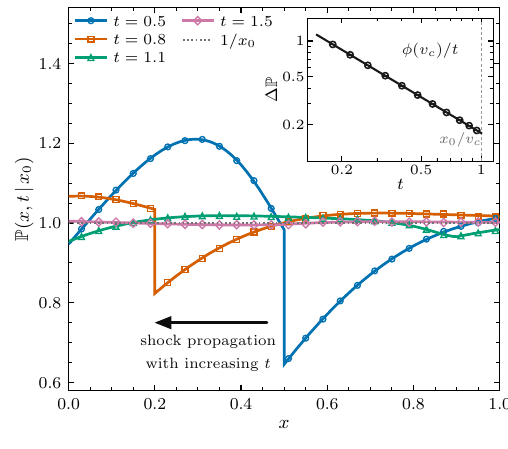}
    \caption{Shock-like transient in the propagator under threshold resetting for a velocity distribution with a lower cutoff, $\phi(v) \propto v^{-3}\exp(-3v_p^2/2v^2)\,\Theta(v-v_c)$. $v_p$ is the mode of this distribution. For $t<x_0/v_c$ the density is discontinuous at the receding front
    $x_d(t)=x_0-v_ct$, drawn as the vertical segment in each curve; the step has height
    $\Delta\mathbb{P}=\phi(v_c)/t$. This jump height is verified with simulation in the inset log-log plot. For $t > x_0/v_c$, there is no discontinuity in the propagator. Solid lines are obtained using numerical inverse Laplace transform of Eq. \eqref{eq:Plap}, and markers denote the Monte-Carlo simulation results. Parameters used: $x_0 = 1, v_c = 1, v_p = 1.6$.} 
    \label{fig:shockTrans}
\end{figure}

The mechanism is transparent from the reset-free propagator. Because of the presence of the cut-off $v_c$, the particle can't be in the region $x > x_d(t)$ at time $t$ in the case when the particle hasn't gone through any resetting, where $x_d(t)$ is defined as follows
\begin{equation}
    x_d(t) = x_0 - v_c t.
\end{equation}
Note that this front propagates towards the origin with time and leaves the domain $[0, x_0]$ after $t = x_0/v_c$.
Since the positions with $x > x_d(t)$ (which would require $v < v_c$) are unreachable in the reset-free case, the probability density drops abruptly to zero for the absorbing boundary propagator
\begin{equation}
\mathbb{G}_+(x,t|x_0)=
\begin{cases}
\dfrac{1}{t}\,\phi\!\Big(\dfrac{x_0-x}{t}\Big), & 0<x\le x_d(t),\\[6pt]
0, & x_d(t)<x \le x_0,
\end{cases}
\label{eq:Gshock}
\end{equation}
where the jump height is defined as
\begin{align}
    \Delta{\mathbb{G}_+}\big|_{x \to x_d(t)} &= \big|\mathbb{G}_+(x_d^+,t|x_0)-\mathbb{G}_+(x_d^-,t|x_0)\big| \nonumber\\
    &=\frac{\phi(v_c)}{t},
    \quad \text{with} \quad t\in\Big(0,\frac{x_0}{v_c}\Big).
\label{eq:jump}
\end{align}
This is the exact jump discontinuity we find in the propagator under TR. It can be explained simply by rewriting Eq. \eqref{eq:renewal} in the last renewal approach \cite{de2020optimization}
\begin{align}\label{eq:PropRecur}
    \mathbb{P}(x, t|x_0) = & \mathbb{G}_{+}(x,t|x_0) + \nonumber \\
    &\sum_{m \ge 1} \int_0^t dt^\prime \, \mathbb{F}_m(t^\prime|x_0)\mathbb{G}_{+}(x, t - t^\prime|x_0),
\end{align}
where $\mathbb{F}_m(t|x_0)$ is the first-passage time distribution for reaching the threshold for the $m$-th time (so that $\mathbb{F}_1 \equiv \mathbb{F}$). The jump discontinuity in $\mathbb{P}$ at $x= x_d(t)$ arises from the first term in Eq. \eqref{eq:PropRecur}, and the convolution term does not contribute to the jump as $\mathbb{G}_+(x, t - t^\prime|x_0)$ has a discontinuity at $x = x_0 - v_c (t -t^\prime)$, but is continuous at $x = x_d(t)$. Consequently, the jump height in the TR propagator is the same as in the $\mathbb{G}_+$ propagator,
\begin{align}\label{Pjump}
    \Delta \mathbb{P}\big|_{x \to x_d(t)} &= \big|\mathbb{P}(x_d^+,t|x_0)-\mathbb{P}(x_d^-,t|x_0)\big|\nonumber\\
    &= \frac{\phi(v_c)}{t},
    \quad \text{with} \quad t\in\Big(0,\frac{x_0}{v_c}\Big).
\end{align}
The jump height depends only on the velocity distribution $\phi(v)$ evaluated at the cutoff $v_c$, and it decays with time. Simulation verification of these results is shown in Fig. \ref{fig:shockTrans}. 

A similar discussion holds in the presence of an upper cutoff, i.e., $\phi(v)$ is defined for $v \in (0, v_m]$. In this case, the jump height becomes $\Delta \mathbb{P} = \phi(v_m)/t$, but now the discontinuity happens at position $x_d^\prime(t) = x_0 - v_m t$. The same goes for $\phi(v)$ with both upper and lower cutoffs, i.e., $v \in [v_c, v_m]$. In this case, discontinuities occur at both $x_d(t)$ and $x^\prime_d(t)$ positions with the jump heights mentioned above.

\section{Deterministic limit: Periodic steady state}\label{sec:deltaFunc}

It is also interesting to ask what happens when there is no randomness in the system, i.e., when the velocity distribution is just a Dirac delta function,
\begin{equation}
    \phi(v)=\delta(v-v_0).
    \label{eq:delta}
\end{equation}
The particle always starts each excursion with the same velocity $v_0$. The propagator under TR follows from Eq. \eqref{eq:Plap}
\begin{equation}
    \widetilde{\mathbb{P}}(x,s|x_0)=\frac{\Theta(x_0-x)}{v_0}\,
    \frac{\exp\!\big[-s(x_0-x)/v_0\big]}{1-\exp\!\big[-sx_0/v_0\big]}.
\label{eq:Pdelta}
\end{equation}
Every pole of Eq.~\eqref{eq:Pdelta} lies on the imaginary $s$-axis, so the
final-value theorem cannot be applied to compute the steady state. By expanding the denominator as a geometric sum, we can compute the inverse Laplace transform. This gives a train of delta peaks
\begin{equation}
    \mathbb{P}(x,t|x_0)=\Theta(x_0-x)\sum_{m\ge0}
    \delta\!\big(x-(x_0-v_0 t+mx_0)\big),
\label{eq:Ptrain}
\end{equation}
a single peak sweeping from $x_0$ to the origin at speed $v_0$ and reappearing at $x_0$ each time $t$ crosses a multiple of $x_0/v_0$. The state is strictly periodic
\begin{equation}
\mathbb{P}(x,t|x_0) = \mathbb{P}(x,t+T|x_0),\qquad T=\frac{x_0}{v_0}=\langle T\rangle,
\end{equation}
with period equal to the deterministic MFPT. The purely imaginary poles in $\widetilde{\mathbb{P}}(x, s|x_0)$ correspond to undamped oscillations in every observable; hence, they are also periodic.

\section{Conclusion}\label{sec:conclude}

In this work, we have investigated a ballistic random walker under threshold resetting
(TR). The walker starts at $x_0$ and moves deterministically towards the threshold at the
origin, with its speed independently redrawn from the distribution $\phi(v)$ after each
reset. The renewal structure of the process allows its dynamics under TR to be
characterized entirely in terms of the corresponding reset-free quantities. Our analysis
of the long-time propagator reveals an interesting universality. Whenever the velocity is
random and the mean first-passage time (MFPT) to the threshold,
$\langle T\rangle = x_0\langle 1/v\rangle$, is finite, the system approaches a uniform
nonequilibrium steady state (NESS) [Eq.~\eqref{eq:uniform}], irrespective of the form of
$\phi(v)$. This universality follows from the mean local time (MLT), which is spatially uniform on
$[0,x_0]$ [Eq.~\eqref{eq:MLT_ball}], so the steady state, being
the ratio of the MLT to the MFPT [Eq.~\eqref{eq:ness-ratio}], is necessarily uniform.

The transients, on the other hand, depend strongly on the velocity distribution. Different
features of $\phi(v)$ map onto distinct singularity structures of the Laplace-space
propagator $\widetilde{\mathbb{P}}(x,s|x_0)$, and hence onto qualitatively different
relaxation behaviors. Most strikingly, the relaxation rate depends on the position $x$, unlike in typical relaxation phenomena, where the rate is usually independent of spatial coordinate. Such a space-dependent rate arises when
$\phi(v)\simeq\mathcal{C}\,v^{-(m+1)}e^{-a/v}$ as $v\to0$ with $m\in\mathbb{N}$. Quite interestingly, the rate can undergo a transition and freezes onto a constant plateau beyond a
critical point $x^\star$, giving rise to a freezing transition.
Equations~\eqref{eq:low-v-tail} and \eqref{eq:ab} together constitute a necessary and
sufficient criterion for this transition -- a representative case was then fully discussed in 
Sec. \ref{sec:solvableExample}. 
We also studied another case when $\phi(v)$ vanishes as a power law when $v \to 0$ and showed that such fluctuations instead preclude any freezing transition and lead to a slow
$t^{-\alpha}$ approach to the steady state. Intriguingly, a lower cutoff $v_c$ in the support of $\phi(v)$ generates a
shock that travels towards the threshold along $x_d(t)=x_0-v_ct$. The discontinuity remains till time $t<x_0/v_c$, and its
height decays exactly as $\phi(v_c)/t$, and an upper cutoff also produces an analogous front.
In the deterministic limit $\phi(v)=\delta(v-v_0)$, the uniform steady state is replaced by a
strictly periodic steady state.

Several promising directions emerge from the present study. 
For an $N$-particle system, it is quite important to study various observables such as spatial correlations, order and gap statistics, and counting statistics in NESS. For a diffusive system under threshold resetting, these observables were extensively studied in Ref. \cite{biroli2024exact}. It would be interesting to extend this study to $N$ particles with the ballistic dynamics studied here.
A further direction is to relax the assumption of ballistic motion with a fixed velocity between successive resets and allow the velocity to change at random times. Depending on the statistics of these run times, this would connect threshold resetting to run-and-tumble dynamics \cite{evans2018run} and Lévy walks \cite{zaburdaev2015levy}, where reversals or changes in speed can drive the walker away from the threshold and raise new questions about the existence and properties of NESS. Beyond these theoretical extensions, our results suggest that the transient density may provide a means of probing the underlying velocity statistics: the shock structure and relaxation law encode information about the upper cutoff and the low-velocity behavior of \(\phi(v)\), respectively. These observations are possibly realizable in controlled resetting experiments \cite{besga2020optimal,faisant2021optimal,tal2020experimental,altshuler2024environmental,kundu2025emulating,paramanick2024uncovering}. More broadly, our findings can help connect threshold-resetting theory with threshold-driven processes in physical and biological systems, including integrate-and-fire neuronal dynamics \cite{burkitt2006review}, and motivate new ways of extracting dynamical information from transient observations.

\section{Acknowledgments}
The numerical calculations reported in this
work were carried out on the Kamet cluster, which is
maintained and supported by the Institute of Mathematical Science’s High-Performance Computing Center. SNM acknowledges support from ANR Grant No. ANR23- CE30-0020-01 EDIPS. AP acknowledges research funding under the scheme
ANRF/ARGM/2025/001623 from ANRF, India and research support from the Department of Science and Technology, India, SERB Start-up Research Grant Number SRG/2022/000080. SNM and AP also acknowledge the International Research Project (IRP) titled ``Classical and quantum dynamics in out of equilibrium systems'' by CNRS, France. RD and AP gratefully acknowledge research support from the Department of Atomic Energy, Government of India via the Apex project ``Modelling Structure \& Dynamics of Soft Matter''. 

\appendix

\section{Finite-time propagator for the \texorpdfstring{$\phi(v) \propto v^{-n}e^{-v_0/v}$}{phin} family}
\label{app:finite}

The simplest way to compute the inverse Laplace transform of Eq. \eqref{eq:Pn} in Sec. \ref{sec:solvableExample} is to use Laplace inversion tables. After expressing $1/[1 - (v_0/(v_0 + sx_0))^{n - 1}]$ in geometric sum, i.e., $\sum_{k \ge 0} (v_0/(v_0 + sx_0))^{k(n - 1)}$, we can rewrite $\widetilde{\mathbb{P}}(x, s|x_0)$ as following
\begin{align}\label{eq:PnSum}
    \widetilde{\mathbb{P}}(x, s|x_0) &= \sum_{k \ge 0}\Theta(x_0 - x) \frac{(n - 1) v_0^{(k + 1)(n -1)}}{(x_0 - x)^n x_0^{k(n - 1)}}\times \nonumber\\
    & \qquad \left[ \frac{1}{(s + \frac{v_0}{x_0 - x})^n (s + \frac{v_0}{x_0})^{k(n -1)}} \right].
\end{align}
The following inverse Laplace transform is known \cite{bateman_1954_mhd23-e0z22}
\begin{align}
    \mathcal{L}^{-1}\left\{ \frac{1}{(s+a)^\alpha (s+b)^\beta} \right\} = e^{-bt} \frac{t^{\alpha+\beta-1}}{\Gamma(\alpha+\beta)}\nonumber\\ \times~_1F_1\big(\alpha; \alpha+\beta; -(a-b)t\big),
\end{align}
where $_1F_1$ is the confluent hypergeometric function\cite{abramowitz2006handbook} (or Kummer's function). By comparing the values $\alpha, \beta, a, b$ for our function in Eq. \eqref{eq:PnSum}, we find the inverse Laplace transform $\mathbb{P}(x, t|x_0)$ in a summation form
\begin{align}\label{eq:1F1sum}
    \mathbb{P}(x,t\vert{}x_0) = \Theta(x_0-x) (n-1) e^{-\frac{v_0}{x_0}t} \sum_{k=0}^{\infty} \frac{v_0^{(k+1)(n-1)}}{(x_0-x)^n x_0^{k(n-1)}}\nonumber\\ \times\frac{t^{(k+1)n-k-1}}{\Gamma\big((k+1)n-k\big)} \, {}_1F_1\left(n; (k+1)n-k; -\frac{v_0 x}{x_0(x_0-x)} t \right).
\end{align}
This result is valid for any $n > 1$, for both integer and non-integer cases. One can also compute this Laplace inversion using the Bromwich integral method.
\begin{equation}
    \mathbb{P}(x, t|x_0) = \frac{1}{2\pi i}\int_{\Gamma_1}ds~e^{s t} \widetilde{\mathbb{P}}(x, s|x_0),
\end{equation}
where $\Gamma_1$ is the line of integration in the complex $s$-plane such that all the singularities of $\widetilde{\mathbb{P}}(x, s|x_0)$ remain on the left side of it (see Fig. \ref{fig:contour}). Because the singularity structure depends on whether $n$ is an integer or non-integer, we discuss each case separately.

\subsection{Integer n}

\begin{figure*}
    \centering
    \includegraphics[width=0.85\linewidth]{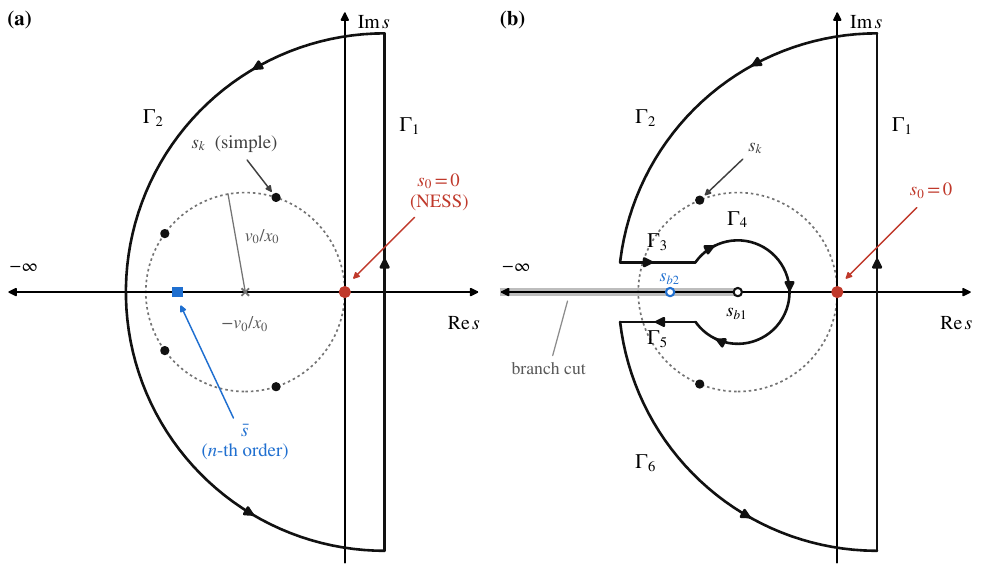}
    \caption{Bromwich contours used to invert $\widetilde{\mathbb{P}}(x,s|x_0)$, traversed
counter-clockwise; $\Gamma_1$ lies to the right of all singularities.
(a) Integer $n$ ($n=6$ taken as an example)[Eq. \eqref{eq:intN}]: the simple poles $s_k$ (filled) lie on the dotted
circle $|s+v_0/x_0|=v_0/x_0$, with $s_0=0$ (red) giving the steady state, and
$\bar{s}=-v_0/(x_0-x)$ (blue) is a space-dependent pole of order $n$. We discuss the exact positioning of these poles in more detail in Sec. \ref{sec:solvableExample} for integer $ n$.
(b) Non-integer $n$ ($n=4.2$ taken as an example)[Eq. \eqref{eq:nonIntres}]: $s_{b1}=-v_0/x_0$ becomes a branch point with a
cut to $-\infty$ (shaded) on which $s_{b2}=-v_0/(x_0-x)$ lies; only the poles
with $|k|<(n-1)/2$ survive on the principal sheet and contribute to the Bromwich integral. These singularities are discussed in Sec. \ref{sec:solvableExample} for non-integer $n$ in the main text and Sec. \ref{app:nonIn} in the appendix.}
    \label{fig:contour}
\end{figure*}

The integrand has an $n$-th-order space-dependent pole at $\bar{s}$ and simple space-independent poles at $s_k$. Exact expressions are given in Eq. \eqref{eq:sbar} and Eq. \eqref{eq:sk} in the main text. Using the Cauchy integral theorem, we can write
\begin{align}\label{eq:intN}
     \frac{1}{2\pi i}\left( \int_{\Gamma_1} + \int_{\Gamma_2}\right) ds \exp(st)\mathbb{\widetilde{P}}(x,s|x_0)\nonumber\\
    =  \sum_{k = 0}^{n-2}\text{Res}(s = s_k) + \text{Res}(s = \bar s),
\end{align}
where the contour lines $\Gamma_1, \Gamma_2$ are shown in Fig. \ref{fig:contour}(a). The contribution due to $\Gamma_2$ vanishes by Jordan's Lemma.
After computing all the residues, we find the full time-dependent expression of the propagator
\begin{align}\label{pdf_velDists}
    \mathbb{P}(x, t|x_0) &= \sum_{k = 0}^{n-2} \exp(s_k t)\frac{v_0^{n-1}(s_k + v_0/x_0)}{(x_0-x)^n (s_k + \frac{v_0}{x_0 - x})^n}\nonumber\\ 
    &+ \exp(\bar{s} t)\sum_{m=0}^{n-1}\frac{h^{(n-1-m)}(\bar{s})}{m! ~(n-1-m)!}t^m,
\end{align}
where $h^{(z)}(s)$ is the $z$-th derivative of $h(s)$ with respect to $s$, and $h(s)$ is defined as follows
\begin{align}
    h(s) = \frac{(n-1) v_0^{n-1}}{(x_0 - x)^n \left(1 - (\frac{v_0}{v_0 + s x_0})^{n-1} \right)}.
\end{align}
The steady-state contribution comes from the pole $s_k = 0$. Both $s_k (k\ne 0) $ and $ s_x$ have a negative real part; they give rise to exponentially decaying relaxation towards the steady state.

\subsection{Non-integer n}\label{app:nonIn}

\begin{figure}[h]
    \centering
    \includegraphics[width=1\linewidth]{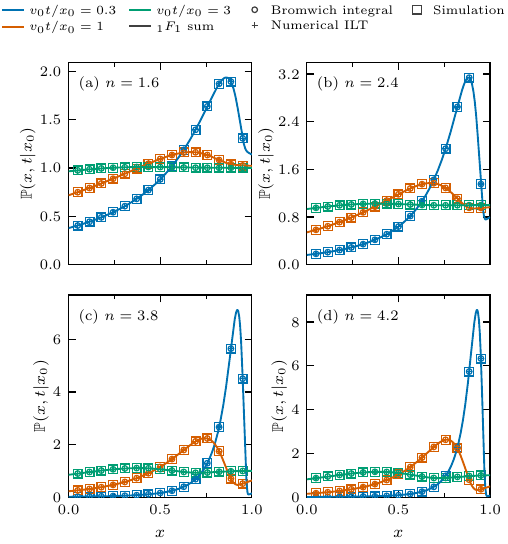}
    \caption{Finite-time propagator $\mathbb{P}(x,t|x_0)$ for non-integer $n$ with
    $\phi(v)$ given in Eq. \eqref{eq:phin}:
    (a) $n=1.6$, (b) $n=2.4$, (c) $n=3.8$, (d) $n=4.2$, at
    $v_0t/x_0=0.3,\,1,\,3$ (blue, orange, green).
    Solid lines are the $_1F_1$ sum using Laplace inversion tables [Eq.~\eqref{eq:1F1sum}], circles the Bromwich-integral result
    [Eq.~\eqref{eq:BromNonInt}], crosses a
    direct numerical Laplace inversion of Eq.~\eqref{eq:Pn}, and squares Monte Carlo simulations. The three analytic and simulation results agree well.
    Parameters: $x_0=1$, $v_0=1$.}
    \label{fig:nonIntVeri}
\end{figure}

For non-integer $n$ cases, the multi-valued terms present in the propagator give rise to branch points instead of full set of poles: a
space-independent branch point at $s_{b1}=-v_0/x_0$ and a space-dependent one at
$s_{b2}=-v_0/(x_0-x)$, with $s_{b2}\le s_{b1}$ for all $x\in[0,x_0]$. The
principal branch point $s_{b1}$ is always bound $s_{b2}$ from right, so the long-time decay
is space-independent and no freezing transition survives for non-integer $n$.
Poles $s_k$ from $(v_0/(v_0+s_k ~x_0))^{n-1}= e^{i2\pi k}$ appear on the principal sheet only when
$|k|<(n-1)/2$ (none for $n<3$, one conjugate pair for $3<n<5$, and so on). Using the Cauchy integral theorem, we write
\begin{align}\label{eq:nonIntres}
    \frac{1}{2\pi i} &\sum_{k=1}^{6} \int_{\Gamma_k} ds\, e^{st}\, \widetilde{\mathbb{P}}(x, s|x_0)
    \nonumber\\
    &= \sum_{k = 0}^{\lfloor (n-1)/2\rfloor} \text{Res}\left[e^{st}\widetilde{\mathbb{P}}(x, s|x_0)\right],
\end{align}
where $\lfloor z\rfloor$ is the greatest integer less or equal to $z$ and the contours are shown in Fig. \ref{fig:contour}(b). The contribution due to contour line $\Gamma_2, \Gamma_4, \Gamma_6$ vanishes\cite{arfken2011mathematical}. $\Gamma_3, \Gamma_5$ contribution arises from the branch cut from $s = s_{b1}$ to $-\infty$.
\begin{equation}
    \frac{1}{2 \pi i} \left(\int_{\Gamma_3} + \int_{\Gamma_5}\right)ds~ e^{s t} ~\widetilde{\mathbb{P}}(x, s|x_0) = - \mathbb{P}_{\text{bp}}(x, t),
\end{equation}
where $-ve$ is chosen on the RHS simply for convenience. After computing the residues and the branch point contribution, we can express $\mathbb{P}(x, t|x_0)$ as follows
\begin{equation}\label{eq:BromNonInt}
\mathbb{P}(x,t|x_0)=\pss(x) + \mathbb{P}_{\mathrm{poles}}(x,t) + \mathbb{P}_{\mathrm{bp}}(x,t),
\end{equation}
where the steady state $\mathbb{P}_{\text{ss}}(x) = 1/x_0$ arises due to the pole at $s = 0$ (or the contribution of the $k = 0$ residue in Eq. \eqref{eq:nonIntres}) and $\mathbb{P}_{\mathrm{poles}}$ is given by the residues of the rest of the poles other than the one at $s = 0$ existing on the principal sheet. The exact expressions of these quantities are given below.
\begin{align}
\mathbb{P}_{\text{poles}}(x,t)=2x_0^{\,n-1}\sum_{k=1}^{\lfloor(n-1)/2\rfloor}
\frac{e^{-\frac{v_0 t}{x_0}(1-\cos\theta_k)}}{R_k(x)^{n}}\nonumber\\
\times\cos\!\Big[\frac{v_0 t}{x_0}\sin\theta_k+n\theta_k-n\varphi_k(x)\Big],
\label{eq:Posc}
\end{align}
where, for the simplicity of the expression, we have defined
\begin{align}
\theta_k &=\frac{2\pi k}{n-1}, \\
\varphi_k(x) &=\tan^{-1}\!\Big[\frac{(x_0-x)\sin\theta_k}
{x+(x_0-x)\cos\theta_k}\Big], \\
R_k(x) &=\sqrt{x^{2}+(x_0-x)^{2}+2x(x_0-x)\cos\theta_k}.
\end{align}
The branch-cut contribution can be computed using $s = s_{b1} + \rho e^{\pm i \pi}$, where $\pm$ denotes the upper ($\Gamma_3$) and lower ($\Gamma_5$) branch-cut lines and $\rho \in [0, \infty)$.
\begin{widetext}
\begin{align}\label{eq:bpnonInt}
    \mathbb{P}_{\text{bp}}(x, t) = -\frac{1}{2\pi i}\int_0^\infty d\rho~ e^{(-v_0/x_0-\rho) t} \left[\widetilde{\mathbb{P}}(x, -\frac{v_0}{x_0} + \rho e^{i\pi} |x_0) - \widetilde{\mathbb{P}}(x, -\frac{v_0}{x_0} + \rho e^{-i\pi} |x_0)\right].
\end{align}
The value of the Laplace space propagator on the integration lines is as follows
\begin{align}\label{eq:onIntLine}
    \widetilde{\mathbb{P}}\Big(x, -\tfrac{v_0}{x_0} + \rho e^{\pm i\pi}\Big|x_0\Big)
    = \frac{(n - 1)v_0^{n - 1}}{(x_0 - x)^{n}}
      \frac{1}{\left[\dfrac{v_0 x}{x_0(x_0 - x)} - \rho\right]^n}
      \frac{(x_0\rho\, e^{\pm i\pi})^{n - 1}}{(x_0\rho\, e^{\pm i\pi})^{n - 1}- v_0^{n - 1}}.
\end{align}
Due to another second branch point at $s_{b2}$, we find a critical point $\rho_c$ defined as
\begin{align}
    \rho_c = \frac{v_0 x}{x_0(x_0 - x)}.
\end{align}
There are two regimes $0 < \rho < \rho_c$ and $\rho \ge \rho_c$, due to the term $[\frac{v_0x}{(x_0- x)x_0} - \rho]^n$ present in the denominator in Eq. \eqref{eq:onIntLine} as it produces an extra phase factor for $\rho > \rho_c$, which makes the discontinuities different between these regimes. For simplicity of notation, let us define
\begin{align}
    &\mathcal{X}(\rho) = \frac{1}{\left|\frac{v_0 x}{x_0(x_0 - x)} - \rho\right|^{n}},\\
    &\mathcal{X}_\pm(\rho) = \mathcal{X}(\rho)\, e^{\mp i n \pi},
    \qquad
    \mathcal{Y}_\pm(\rho) = \frac{(x_0\rho)^{n-1}\, e^{\pm i\pi(n-1)}}{(x_0\rho)^{n-1}\, e^{\pm i\pi(n-1)} - v_0^{\,n-1}}.
\end{align}
By using the above-mentioned variables, we can express the different discontinuities in both regions as below
\begin{align}
    \widetilde{\mathbb{P}}(x, -\frac{v_0}{x_0} + \rho e^{i\pi} |x_0) - \widetilde{\mathbb{P}}(x, -\frac{v_0}{x_0} + \rho e^{-i\pi} |x_0) = \frac{(n - 1)v_0^{n - 1}}{(x_0 - x)^{n}} \times
    \begin{cases}
        \mathcal{X}(\rho)\left[\mathcal{Y}_+ (\rho) - \mathcal{Y}_- (\rho)\right], & \rho < \rho_c, \\
        \mathcal{X}_+(\rho)\mathcal{Y}_+(\rho) - \mathcal{X}_-(\rho)\mathcal{Y}_-(\rho), & \rho \ge \rho_c.
    \end{cases}
\end{align}
After doing a few steps, we find the exact expression of the above quantities,
\begin{align}
    &\mathcal{Y}_+ (\rho)- \mathcal{Y}_-(\rho) = -\,i\,\frac{2 (x_0 v_0 \rho)^{n - 1} \sin[\pi(n - 1)]}{\mathcal{D}(\rho)},\\
    &\mathcal{X}_+(\rho)\mathcal{Y}_+(\rho) - \mathcal{X}_-(\rho)\mathcal{Y}_-(\rho)
    = \mathcal{X}(\rho)\left(e^{-in\pi}\mathcal{Y}_+(\rho) - e^{in\pi}\mathcal{Y}_-(\rho)\right)
    = i\,\mathcal{X}(\rho)\,\frac{2 (x_0 \rho)^{2(n - 1)} \sin[\pi(n - 1)]}{\mathcal{D}(\rho)}, \\
    &\mathcal{D}(\rho) = (x_0 \rho)^{2(n -1)} - 2 (x_0 v_0 \rho)^{n - 1}\cos[\pi(n - 1)] + v_0^{2(n-1)}.
\end{align}
Using the above results, the final expression of the branch-cut integral reads
\begin{equation}\label{eq:bpNumInt}
\begin{aligned}
\mathbb{P}_{\mathrm{bp}}(x,t)
={}&
\frac{(n-1)v_0^{n-1}\sin\!\bigl(\pi(n-1)\bigr)}
{\pi (x_0-x)^n}
e^{-v_0t/x_0}
\Biggl[
 v_0^{n-1}
\int_0^{\rho_c}
\frac{e^{-\rho t}(x_0\rho)^{n-1}}
{\mathcal{D}(\rho)(\rho_c-\rho)^n}\,d\rho
-\int_{\rho_c}^{\infty}
\frac{e^{-\rho t}(x_0\rho)^{2(n-1)}}
{\mathcal{D}(\rho)(\rho-\rho_c)^n}\,d\rho
\Biggr].
\end{aligned}
\end{equation}
Exactly at $\rho = \rho_c$, the above integral diverges for any $n > 1$. We can overcome this problem by computing the finite part integral, which is nothing but computing the above integral without considering the point $\rho = \rho_c$. For a detailed discussion and comparison with the Cauchy principal value integrals, see Ref. \cite{monegato2009definitions}. A finite part integral can be defined in this way
\begin{align}
    \ddashint_a^b dz \frac{f(z)}{(z - a)^{\alpha + 1}} = \frac{1}{\alpha} \frac{1}{da}\int_a^bdz \frac{f(z)}{(z - a)^\alpha}, \qquad 0 < \alpha < 1,
\end{align}
where $f(z)$ is a differentiable function at $z = a$. The RHS integral is finite. In our case, $n > 1$ can be any non-integer value, which may require multiple derivatives with respect to $\rho_c$; e.g., the above operation is applicable when $1 < n < 2$, where we can directly compare $\alpha + 1$ with $n$. For an arbitrary $n$, we define $\bar{n} = \lfloor n \rfloor$ and $n_0 = n - \bar{n} \in (0, 1)$. Using these, we can compute the finite part integral of Eq. \eqref{eq:bpNumInt} as below
\begin{equation}\label{eq:bpNumIntFP}
\begin{aligned}
\mathbb{P}_{\mathrm{bp}}(x,t)
={}
\frac{(n-1)v_0^{n-1}\sin\!\bigl(\pi(n-1)\bigr)}
{\pi (x_0-x)^n}
e^{-v_0t/x_0}
&\Biggl[
 v_0^{n-1}(-1)^{\bar{n}} \frac{\Gamma(n_0)}{\Gamma(n)}\left(\frac{d}{d\rho_c} \right)^{\bar{n}}
\int_0^{\rho_c}
\frac{e^{-\rho t}(x_0\rho)^{n-1}}
{\mathcal{D}(\rho)(\rho_c-\rho)^{n_0}}\,d\rho \\
& \qquad-\frac{\Gamma(n_0)}{\Gamma(n)}\left(\frac{d}{d\rho_c} \right)^{\bar{n}}\int_{\rho_c}^{\infty}
\frac{e^{-\rho t}(x_0\rho)^{2(n-1)}}
{\mathcal{D}(\rho)(\rho-\rho_c)^{n_0}}\,d\rho
\Biggr],
\end{aligned}
\end{equation}
where the function $e^{-\rho t}\rho^{n - 1}/\mathcal{D}(\rho)$ is $\bar{n}$ times differentiable at $\rho = \rho_c$. Fig. \ref{fig:nonIntVeri} shows a verification plot of the different analytical results for the inverse Laplace transform $\mathbb{P}(x, t|x_0)$ for non-integer $n$ values, along with simulation.
\end{widetext}

\section{Exactly solvable n = 3 example:}\label{app:n=3 case}

To expose the physics of the exponential relaxation and freezing transition, we adopt the velocity
distribution
\begin{equation}
\phi(v)=\frac{v_0^{2}}{v^{3}}\,e^{-v_0/v},\qquad v\ge0,
\label{eq:phi3}
\end{equation}
where $v_0$ is a characteristic speed. Due to the presence of an essential singularity in the small-$v$ limit, the MFPT to find the threshold is finite, and consequently there exists a steady state.
After carrying out the Laplace inversion using the Bromwich integral method mentioned in Appendix~\ref{app:finite}, the propagator under TR can be expressed as follows
\begin{equation}
    \mathbb{P}(x,t|x_0)=\frac{1}{x_0}\,\mathcal{F}\left(\bar x = \frac{x}{x_0},\bar t = \frac{v_0 t}{x_0}\right)\,\Theta(x_0-x),
    \label{eq:Pexact}
\end{equation}
with the scale function defined as
\begin{align}
&\mathcal{F}(\bar x,\bar t)= 1 
+\mathcal{G}_1(\bar x, \bar t)~e^{-\bar t/(1-\bar x)}+ \mathcal{G}_2(\bar x, \bar t)~e^{-2\bar t},
\label{eq:scalefun}
\end{align}
where
\begin{align}
    \mathcal{G}_1(\bar x,\bar t) &= \frac{4\bar t\,\bar x}{(2\bar x-1)^2} +\frac{\bar t^{2}\bar x^{2}}{(1-\bar x)^3(2\bar x-1)}\nonumber\\
    & \qquad +\frac{2(1-\bar x)(4\bar x^{2}-2\bar x+1)}{(2\bar x-1)^3},\\
    \mathcal{G}_2(\bar x,\bar t) &= -\frac{1}{(2\bar x-1)^3}.
\end{align}
Although Eq.~\eqref{eq:scalefun} appears singular at $\bar x=1/2$, the
singularities cancel, and the limit is finite,
\begin{equation}
\lim_{\bar x\to1/2}\mathcal{F}(\bar x,\bar t)
=1-e^{-2\bar t}\Big(1+2\bar t-6\bar t^{2}+\tfrac{4}{3}\bar t^{3}\Big),
\label{eq:half}
\end{equation}
so the density is smooth throughout $[0,x_0]$. As $\bar t\to\infty$,
$\mathcal{F}\to1$ at every $\bar x$, recovering the universal uniform steady
state~\eqref{eq:uniform}. At any fixed $\bar x$, the relaxation of
$\mathcal{F}(\bar x,\bar t)$ is non-monotonic in time---the density overshoots
and undershoots its stationary value before settling---reflecting the interplay
of the several decay modes present in Eq.~\eqref{eq:scalefun}. These non-monotonic relaxations of different spatial points are shown in Fig. \ref{fig:n=3 case}(e).

\subsection{Freezing transition in the relaxation rate}
\label{sec:freeze-example}
 
At long times, the slowest decay mode dominates, and the density relaxes as
\begin{equation}
\mathbb{P}(x,t|x_0)\simeq\pss(x) + \mathcal{H}(x,t)\,e^{-\lambda(x)t},
\label{eq:relax}
\end{equation}
where $\mathbb{P}_{\text{ss}}(x) = \Theta(x_0 -x)/x_0$ is the steady state, $\mathcal{H}(x,t)$ is algebraic in $x,t$ and $\lambda(x)$ is the relaxation
rate. 
Reading off the dominant exponential in Eq.~\eqref{eq:scalefun}, we find
\begin{equation}
\lambda(x)=
\begin{cases}
\dfrac{v_0}{x_0-x}, & 0\le x<\dfrac{x_0}{2},\\[10pt]
\dfrac{2v_0}{x_0}, & \dfrac{x_0}{2}\le x\le x_0.
\end{cases}
\label{eq:lambda-example}
\end{equation}
The relaxation is thus spatially heterogeneous. In one part it is space-dependent, and in the rest it is space-independent. The
crossover point $x^\star=x_0/2$, where the rate stops depending on position and
freezes to a plateau value, is called a
freezing transition-of the same type identified in the barrier
crossing rate of a diffusing particle~\cite{sabhapandit2020freezing}. Its mathematical
origin lies in the competition between two families of singularities of
$\widetilde{\mathbb{P}}(x,s|x_0)$, which we analyze in full generality in Sec.~\ref{sec:FreezTrans} in the main text.

\subsection{Simulation verification in Fig. \ref{fig:n=3 case}(a), (b)}

For verification purposes in Fig. \ref{fig:n=3 case}, we represented the propagator as
\begin{align}
    \mathbb{P}(x, t|x_0) = \mathbb{P}_{\text{ss}}(x) + h_1(x, t) e^{-v_0 t/(x_0 - x)}\nonumber\\
    + h_2 (x, t) e^{-2v_0 t/x_0},
\end{align}
where $h_1$ and $h_2$ can be defined using the dimensionless coefficients $\mathcal{G}_1$ and $\mathcal{G}_2$ as below
\begin{align}
    h_1(x, t) &= \frac{1}{x_0}\mathcal{G}_1\left(\frac{x}{x_0}, \frac{v_0 t}{x_0}\right),\\ h_2(x, t) &= \frac{1}{x_0}\mathcal{G}_2\left(\frac{x}{x_0}, \frac{v_0 t}{x_0}\right).
\end{align}

\bibliographystyle{unsrt}
\bibliography{references}

\end{document}